\documentstyle[12pt] {article}                                                  
\begin{document}
\textheight=23.5 cm                                                              
\textwidth=17.5 cm                                                              
\voffset=-1.5 cm                                                                
\hoffset=-1.5 cm   
%
\def\ie{{\it i.e. \/}}
\def\eg{{\it e.g. \/}}
\def\cf{{\it c.f. \/}}
\def\viz{{\it viz.\/}}
\def\cap{\caption}
\def\cline{\centerline}
\def\half{\scriptstyle{\frac{1}{2}}} 
\def\halft{\textstyle{\frac{1}{2}}} 
\def\osqrt{\textstyle{\frac{1}{\sqrt2}}} 
\def\lsqrt{\textstyle{\frac{\l}{\sqrt2}}} 
\def\phalf{\textstyle{\frac{\pi}{2}}}  
\def\nd{\noindent}
\def\nn{\nonumber}
\def\ad{a^\dagger}
\def\ab{\bar{\alpha}}
\def\nub{\bar{\nu}}
\def\hi{\chi_{klm}}
\def\udp{U_{\lambda}^{\dagger}}
\def\udm{U_{-\lambda}^{\dagger}}
\def\utp{\tilde{U}_{\lambda}}
\def\utm{\tilde{U}_{-\lambda}}
\def\up{U_{\lambda}}
\def\um{U_{-\lambda}}
\def\so{\sigma_{1}}
\def\st{\sigma_{2}}
\def\sth{\sigma^{z}}
\def\sp{\sigma^{+}}
\def\sm{\sigma^{-}}
\newcommand{\fscr}[2]{\scriptstyle \frac{#1}{#2}} 
\newcommand{\bra}[1]{\left<#1\right|}
\newcommand{\ket}[1]{\left|#1\right>} 
\newcommand{\braket}[1]{\left<#1\right>}
\newcommand{\inner}[2]{\left<#1|#2\right>}
\newcommand{\sand}[3]{\left<#1|#2|#3\right>}
\newcommand{\proj}[2]{\left|#1\left>\right<#2\right|}
\newcommand{\absqr}[1]{{\left|#1\right|}^2}
\newcommand{\abs}[1]{\left|#1\right|}
\newcommand{\lag}[2]{L_{#1}^{#2}(4\l^{2})}
\newcommand{\dbl}[2]{\rm#1\hskip-.5em \rm#2}
\newcommand{\mes}[1]{d\mu(#1)}
\newcommand{\pl}[1]{\partial_{#1}}
\newcommand{\mat}[4]{\left(\begin{array}{cc}
                           #1 & #2 \\
                           #3 & #4
                          \end{array}\right)}
\newcommand{\col}[2]{\left( \begin{array}{c}
                                 #1 \\
                                 #2
                            \end{array} \right)}
\def\a{\alpha}
\def\b{\beta}
\def\g{\gamma}
\def\d{\delta}
\def\e{\epsilon}
\def\z{\zeta}
\def\th{\theta}
\def\f{\phi}
\def\l{\lambda}
\def\m{\mu}
\def\p{\pi}
\def\om{\omega}
\def\D{\Delta}
\def\zb{\bar{z}}
\newcommand{\be}{\begin{equation}}                                              
\newcommand{\ee}{\end{equation}}                                                
\newcommand{\ba}{\begin{array}}  
\newcommand{\ea}{\end{array}}                                                
\newcommand{\bea}{\begin{eqnarray}}                                              
\newcommand{\eea}{\end{eqnarray}}                                                
\newcommand{\beann}{\begin{eqnarray*}}                                         
\newcommand{\eeann}{\end{eqnarray*}}                                                
\newcommand{\bfg}{\begin{figure}}                                                
\newcommand{\efg}{\end{figure}}                                                
%
%
\begin{titlepage}                                                               
\vskip 0.5 cm          

\begin{center}                                                                  
{\Large\bf $q$-Symmetries in DNLS-AL chains and exact solutions of 
quantum dimers}

\vskip 0.5 cm                                                                   
                                                                                
{\bf Demosthenes Ellinas $^1$}\renewcommand{\thefootnote}{\sharp}\footnote{     
Email: {\tt ellinas@science.tuc.gr}}
and \
{\bf Panagiotis Maniadis  $^2$ }
\renewcommand{\thefootnote}{\diamond}\footnote{
Email: {\tt maniadis@physics.uch.gr}}\\
\vskip 0.5cm
$^1$ Department of Sciences \\
Technical Univercity of Crete\\
 GR - 73 100 Chania  Crete Greece.\\
\vskip 0.2cm
$^2$ 
Department of Physics University of Crete \\
and Research Center of Crete (FORTH)\\
P.O.Box 2208, Heraklion GR - 71 003 Crete Greece.\\
\vskip 1.0 cm                                                                   
\end{center}                                                                    
\begin{abstract}
Dynamical symmetries of Hamiltonians quantized  models of 
discrete non-linear Schr\"odinger chain (DNLS) and of 
Ablowitz-Ladik chain (AL) are studied. It is shown that for $n$-sites
the dynamical algebra of DNLS Hamilton operator is given by the 
$su(n)$ algebra, while the respective symmetry for the AL case is 
the quantum algebra $su_q(n)$. The $q$-deformation of the dynamical
symmetry in the AL model is due to the non-canonical oscillator-like
structure of the raising and lowering operators at each site.

Invariants of motions are found 
in terms of Casimir central elements of $su(n)$
and $su_q(n)$ algebra generators, for the DNLS and QAL cases respectively. 
Utilizing the representation theory of the symmetry algebras 
we specialize to the $n=2$ quantum dimer case and formulate the 
eigenvalue problem of each dimer as a non-linear ($q$)-spin 
model. Analytic investigations of the ensuing 
three-term non-linear recurrence relations are carried out and the respective
orthonormal and complete eigenvector bases are determined.

\nd The quantum manifestation of the classical self-trapping in 
the QDNLS-dimer and its absence in the QAL-dimer, is analysed
by studying the asymptotic attraction and repulsion respectively,
of the energy levels versus the strength of non-linearity. Our 
treatment predicts for the QDNLS-dimer,
a phase-transition like behaviour in the rate of change of
the logarithm of eigenenergy differences, for values of the non-linearity 
parameter near the classical bifurcation point.
\end{abstract}

\vskip 1.0 cm
\end{titlepage}
\pagenumbering{arabic}
\section{Introduction}

The Discrete Nonlinear Schr\"odinger 
and the Ablowitz-Ladik chain models have generated intense interest
during the last decade to physicists and mathematicians.  Both systems 
consist of coupled sets of ordinary differential equations,  the former has 
interesting physical motivations while the latter has attractive mathematical 
properties.  The property of DNLS that makes it physically appealing is that of
selftrapping~\cite{KC,ELS,BES}, \viz the existence of broken symmetry localized states induced
by nonlinearity.  This property  
 makes
DNLS an interesting model for discrete solitons or polarons. 
On the other hand  the AL
equation~\cite{AL},
even though 
not physically as transparent as DNLS has the unique
mathematical property of intergrability, as a result many of its properties can be
investigated analytically\cite{Ger}.  

An issue that already has been
addressed by various authors (see \eg~\cite{BES,ism,Ger,EJC}),
 is that of the quantum properties of these two
equations.  In particular the question of the quantum manifestation of selftrapping
and in what precise way the quantum versions of DNLS and AL differ. This is 
the question
we also address in the present article through the use of novel techniques motivated
from the area of quantum groups~\cite{CP}.  In particular, we develop a systematic
algebraic scheme for the Quantum DNLS (QDNLS) and Quantum AL (QAL) model
 that uses $q$-symmetries. This includes the finding of the dynamical 
 symmetry algebra for each model and the determination 
 of the set of independent constants of the motion for the respective 
 Hamiltonians. It is achieved by using a canonical boson realization for the
 $su(n)$ algebra, in the case of the $n$-site open QDNLS chain. The chain 
 possesses then as invariants the set of $n-1$ independent Casimir operators 
 of the algebra, the eigenvectors of which will determine the 
 linear space of the unitary time
 evolution of the model. These considerations are further extended to the
 case of the QAL chain model. In this case we first show 
 that the quantum dynamical variable at each chain site is a particular case of the
 so called $q$-oscillator~\cite{Mac} (see~\cite{Sal} for related considerations). 
 Then  we invoke a $q$-oscillator realization
 of the $q$-deformed (quantum) $su(n)$ algebra which lead us to determine
 the invariants
 of motion of the QAL Hamiltonian that are given in terms of the quantum 
 Casimir elements. 
 
 In the second part of the work we turn our attention 
 to the special case of
  two-site (dimer) DNLS and QAL models where the preceding 
  general algebraic scheme is now applied. In terms of the representation 
  theory of the respective dynamical algebras of the two dimers we 
  formulate the eigenvalue-eigenvector problem and provide analytic solution
  in each case. Our aim is to use the obtained solutions for a cooperative 
  study of the spectral features of the two dimers from the perspective 
  of the phenomenon of seltrapping.
 In particular,
we investigate the energy level structure, the precise nature of which determines
the relative localization-delocalization transition of the quantum excitation for 
values of the non-linearity parameters near the  
bifurcation of the classical DNLS dimer. The plan of the paper is as follows:
In the following section we discuss the quantum $q$-symmetries of QDNLS and QAL, 
in section 3 we provide the corresponding dimer exact solutions and finally
in section 4
we give the numerical results with a physical interpretation of the
resulting plots along with conclusions and future prospects 
of our work.

\section{ Quantum symmetries }
\subsection{\it The QDNLS model}
In this chapter we will start the investigation of the dynamical 
symmetries of the quantum DNLS (QDNLS) and the quantum AL (QAL) chain.
The two models are quantized versions of their classical analogues.  
The quantization procedure is the usual naive method of substituting 
the classical complex amplitudes by
canonical (and non-canonical as in the AL case)
boson operators, 
following the normal ordering rule, and the Poisson brackets by commutators
\cite{W1}.

We consider first the Hamiltonian of the QDNLS $n$-site open chain,
\be
H_{DNLS}=\sum _{i=1}^n \left\{ -\epsilon (a_{i+1} +a_{i-1})a_i^\dagger
-{\gamma \over{2}}{(a_i^\dagger  a_i) }^2 \right \}
\label{hdnls}
\ee
with the canonical commutation relations
\bea
  \left[a_i,a_j ^\dagger \right] &=& \delta _{ij}\;, \nn \\
  \left[ N _i , a _j \right]&=& -\delta _{ij} a _i \;,\nn \\
  \left[N _i , a_j ^\dagger \right]&=& \delta _{ij} a_i ^\dagger \;,
\label{ccr}
\eea
among the boson sites.

\nd To study the algebraic symmetry of this Hamiltonian we consider the defining 
relations of the $su(n)$ algebra generators, in the  
Cartan-Chevalley basis \cite{Gil}
$\left\{ e_i,f_i,h_i \right\} _{i=1} ^{n-1}$, which read
\bea
\label{sun}
\left[ h_i,h_j\right] &=& 0 \\
\left[ h_i,e_j\right] &=& {1\over{2}}\a_{ij}e_j \;, \nn \\
\left[ h_i,f_i\right] &=& -{1\over{2}}\a_{ij}f_j \;, \nn \\
\left[ e_i,f_i\right] &=& 2h_i\delta_{ij} \;, \nn 
\eea
\bea
\sum_{r,s} (-1)^r \left[ \begin{array}{c}
                         1-a_{ij}\\
			 r
			 \end{array}\right]
			 e_i^r e_j e_i^s &=& 0 \;, \quad i\neq j ,\nn \\
\sum_{r,s} (-1)^r \left[ \begin{array}{c}
                         1-a_{ij}\\
			 r
			 \end{array}\right]
			 f_i^r f_j f_i^s &=& 0 \;, \quad i\neq j 
\label{chev}
\eea

\nd where $r,s$ are non-negative integers constrained by $r+s = 1- \a_{ij}$.
\nd Notice that the last two equations are the Serre relations which involve
cubic products of the $n-1$ generators, in terms of these the remaining
$su(n)$ generators are defined, until the total number of the $n^2 -1$
basis elements is completed. Also the Cartan matrix elements which occur as
structure constants in eq.(\ref{chev}), are defined as
$\a _{ij}=2\delta _{ij}-\delta _{i,j+1} -\delta _{i,j-1}$.
Finally in this algebra basis the raising and lowering generators 
are  
conjugate \ie
$e_i^\dagger =f_i$ and the 
  Cartan subalgebra generators are self-adjoint \viz
\ $h_i^\dagger =h_i$, $i=1,...,n-1$.

\nd Then we invoke the bosonic realization of the $su(n)$ algebra given in terms of 
the set of $n$-bosons of eq.(\ref{ccr}),\cite{barut,Gil},
\bea
e_i &=& a_i ^\dagger a_{i+1} \nn \;,\\
f_i &=& a_{i+1} ^\dagger a_i \nn \;, \\
h_i &=& {1\over{2}} (N_i -N_{i+1})\;.
\label{bosrel}
\eea

\nd Employing the central element $h=N_1+N_2+...+N_n$ (total 
energy operator), we express the number operator at each site 
in the form 
\be
N_i={1\over{2}} \sum _{j=1} ^{n-1} (\Omega ^{-1} ) _{ij} h_i 
                                  +(\Omega ^{-1} ) _{in} h ,
\label{nu}
\ee
where
\be
\Omega =\left[ \begin{array}{c c c c c c}
    1& -1& 0& &...&0\\
    0&  1 &-1&0&...&0\\
    . & & & & & \\
    . & & & & & \\
    . & & & & & -1\\
    1 &1 &... & &1 & 1
    \end{array}
    \right]
 \ee
is an $n \times n $ invertible matrix.

By virtue of the bosonization of the generators the QDNLS Hamiltonian can now 
be embedded in the $su(n)$ algebra 
\be
H_{DNLS}=\sum _{i=1}^n
    \left\{ -\epsilon (e_i + f_i) -
{\gamma \over{2}}\left[{1\over{2}}
\sum _{j=1} ^{n-1} (\Omega ^{-1})_{ij}h_j\right] ^2
- \gamma  h \sum _{j=1} ^{n-1} (\Omega ^{-1})_{ij}h_j 
- {\gamma \over{2}} (\Omega ^{-1})_{in}^{2}h^2 \right\} \;.
\ee

 This dynamical symmetry in turn implies that all central (Casimir)
 elements of the $su(n)$ algebra commute with $H_{DNLS}$,
 and therefore become constants of motion. The number of
 these independent invariants equals the rank of the
 $su(n)$ which is $n-1$. From the various forms available of
 the Casimir operators in terms of the $su(n)$ generators
 \cite{barut} we will use one that facilitates the comparison
 with the corresponding $q$-deformed Casimirs that will be
 discussed next for the case of QAL chain. To this end referring
 to eq.(\ref{sun}), we introduce the generators             
 \bea
 h_{a} &\equiv& \e_{a}-\e_{a +1} \;\; \e_{1}+\e_{2}+\cdots +\e_{n}=0,\nn \\
 E_{ab} &\equiv& \left\{
  \begin{array}{ll}
 E_{ac}E_{cb}- E_{cb}E_{ac} &  a< c< b \;, {\rm no \ summation}\\
 e_{a}& a+1=b\;,
 \end{array}
 \right.
 \eea

 \nd and $E_{ab}=0$, for $a=b$, $a>b$. Also $F_{ba}\equiv E_{ab}^\dagger$,
 $e_a ^\dagger = e_a$, and the range of values for all indices is from
 $1$ to $n$, while the index $c$ above is arbitrary provided that
 $a<c<b$. In terms of these generators the desired Casimir invariants
 are,
 \be
 C_{2p}\equiv(M^p)_{aa},
 \ee

 \nd where the matrix $M$ is defined by $M\equiv E_{ac}F_{cb}$, and
 recursively the higher powers are given be $(M^{p+1})_{ab}=
 (M^{p})_{ac}M_{cb}$. An alternative set of Casimir operators is defined
 by
 \be
 C_{2p}^{'}\equiv(N^{p})_{aa}
 \ee

 \nd with $N_{ab}=F_{ac}E_{cb}$. The above expressions provide two
 sets of Casimir operators of any degree for the algebra $su(n)$,
 \ie $[H_{DNLS},C_{2p}]=[H_{DNLS},C_{2p}^{'}]=0$. The quantum dynamics
 generated by the Hamiltonian $H_{DNLS}$ is constrained by the existence
 of these invariances. The state vector of the system evolves in the
 vector space spanned by the eigenvectors of the set of Casimir operators
 that correspond to a given set of eigenvalues. Both the eigenvalues and
 the set of their respective eigenvectors are labeled by a minimal set
 of indices the so called Casimir indices, the number of which equals
 the rank of the dynamical algebra. This is the standard scheme of
 dynamical symmetries for quantum Hamiltonian systems
 (see \eg \cite{barut}). It can be utilized to determined the dynamics
 of the general DNLS chain and shortly it will be used in the
 DNLS dimer case.

As the spectrum generating algebra of the model, the $su(n)$ 
algebra, provides means for the 
solution of the eigenvalue problem. This is particularly so in the case 
of two sites to which we now turn. Renaming the generators $e_1=J_+$, 
$f_1=J_-$, $h_1=J_0$, in terms of the more usual $su(2)$ spin algebra 
notation,
the quantum dimer Hamiltonian after some rescaling and a shift by a 
constant term, reads (for the dynamics of the wavefunction zeros of this 
Hamiltonian see~\cite{demkov}),
\be 
\label{hqd}
H_{QD}=J_+ +J_- +{\gamma \over{2}} J_0 ^2 \;.
\ee

\nd This form of $H_{QD}$ is particularly suitable for the study of its 
eigenvalue problem which will be taken up in the next chapter.

\subsection{\it The QAL model}
We turn now to the AL discretization of the continues NLS equation \cite{AL}.
This is  a Hamiltonian model with Hamiltonian operator
\be
H_{AL}=-\sum _{i=1} ^n \left \{ b_i ^\dagger (b_{i+1} +b_{i-1} )
       -2{\ln(1+{\gamma \over{2}}b_i ^\dagger b_i ) 
       \over{\ln(1+{\gamma \over{2}} ) }} 
       \right\}\;.
\label{hAL}
\ee
The model employs a set of non-canonical oscillators \cite{Ger} with 
commutation relations
\bea
\left[b_i,b_j ^\dagger \right]&=&(1+{\gamma \over{2}}b_i ^\dagger b_i ) 
             \delta _{ij}\;, \nn\\
\left[N_i,b_j  \right]&=& -b_i \delta _{ij} \;, \nn \\
\left[N_i,b_j ^\dagger \right]&=&b_i ^\dagger \delta _{ij} \;.
\label{nonCCR}
\eea

To demonstrate the quantum group symmetry of the QAL Hamiltonian we need 
first to show the relation of the non-canonical AL-oscillator to the so called
$q$-oscillator (quantum Heisenberg algebra) \cite{Mac} and subsequently to 
consider the $q$-bosonization of the quantum group $su_q (n)$ .
We start with the abstract $q$-oscillator
algebra generated by elements $a$, $a^\dagger $, and $N$ 
with commutation relations~\cite{Mac}
\be
aa^\dagger -qa^\dagger a =q ^{-N}\;, \nn
\ee
\be
\left[ N,a ^\dagger \right]=a ^\dagger \ \ ,\ \ \left[ N,a \right]=-a \;.
\label{qosc}
\ee
The parameter $q$ is the so called deformation parameter and in our case
is taken to be a non-negative real number. The $q$-oscillator is so defined 
that in the so called classical limit when $q\rightarrow 1$, it reduces
to the standard quantum mechanical oscillator. 
This $q$-deformed algebra possesses a nontrivial central element
\be
C=q^{-N} ([N]-a^\dagger a)\;,
\ee
where $[N]=(q^N -q^{-N})/(q-q^{-1})$, which implies that
\be
a^\dagger a=[N]-q^{N} C\;.
\label{Cequ}
\ee
If we introduce the generators 
\be
b^\dagger = a^\dagger q^{-N/2}\ \  {\rm and }
\ \ b=q^{-N/2}a ,
\label{trans}
\ee

\nd we express the defining relations 
in terms of commutators \ie
\be
\left[b,b^\dagger \right]=q ^{-2N}\ ,\ \ 
\left[N,b^\dagger \right]=b^\dagger \ ,\ \ 
\left[N,b \right]=b \;.
\ee
Then eq.(\ref{Cequ}) is equivalent to 
\be
b^\dagger b=q^{-N+1}[N]-qC\;,
\ee
which amounts to 
\be
q^{-2N}=1-qC-(1-q^{-2})b^\dagger b\;.
\ee
To correctly identify the AL-oscillator we make the choice $C=0$,
$q={1\over { \sqrt{1+{\gamma \over{2}} }}}$, which imply the 
relations
\be
\left[ b,b^\dagger \right] =1+{\gamma \over{2}}b^\dagger b
\ee

\nd and
\be
N=\frac{\ln(1+{\gamma \over{2}} b ^\dagger b)}{\ln(1+{\gamma \over{2}})}\;,
\label{numb}
\ee

\nd for the $q$-Heisenberg oscillator. From the representation theory
of the abstract $q$-oscillator algebra \cite{Kulish}, we know that it possesses 
a family of irreducible infinite dimensional representations (irrep),
with various operator properties assigned to its generators. However the choice
$C=0$, singles out a representation for which the number operator $N$,
 is positive definite. This Fock-space matrix irrep of the algebra
constructed in the Hilbert space
${\cal H}_F$, spanned by the basis vector
$\{|n>\} _{n=0} ^{\infty}$,  makes the $H_{AL}$ a proper energy operator
and provides the following matrix realization for its elements: 
\be
N|n>=n|n> \;,
\ee
\be
b|n>=\sqrt{\{n\}}|n-1> \;,
\ee
\be
b^\dagger |n>=\sqrt{\{n+1 \}}|n+1> \;,
\ee

\nd with $b|0>=0$ and $\{n\}={q^n-1 \over{q-1}}$.

We proceed now showing the quantum group symmetry of the QAL-chain. First 
let us recall the defining relations of the quantum $su_q(n)$ algebra 
according to the Jimbo-Drin'feld scheme, in the 
Cartan-Chevalley basis 
$\left \{ e_i,f_i,k_i \equiv q^{h_i} \right \} _{i=1} ^{n-1}$, they are 
\cite{CP}
\bea
k_i k_i^{-1} &=& k_i^{-1} k_i =1 \;, \nn \\
k_i k_j &=& k_j k_i\;, \nn \\
k_i e_j k_i^{-1} &=& q^{{1\over{2}}\a_{ij}} e_j\;, \nn \\
k_i f_j k_i^{-1} &=& q^{- {1\over{2}}\a_{ij}} f_j\;, \nn \\
\left[ e_i,f_i\right] &=& [2 k_i]\delta_{ij} \;, \\
\eea

\nd together with the $q$-Serre relations
\bea
\sum_{r,s} (-1)^r \left[ \begin{array}{c}
                         1-\a _{ij}\\
			 r
			 \end{array}\right]_q
			 e_i^r e_j e_i^s &=& 0 \;, \quad i\neq j ,\nn \\
\sum_{r,s} (-1)^r \left[ \begin{array}{c}
                         1-\a _{ij}\\
			 r
			 \end{array}\right]_q
			 f_i^r f_j f_i^s &=& 0 \;, \quad i\neq j .
\eea

\nd The last ones involve the so called $q$-binomial coefficient
\cite{Ext} defined in terms of $q$-factoral,
$[m]!=[1][2]...[m]$ by
\be
\left[
\begin{array}{c}
m\\
n
\end{array}
\right] _q = { [m]! \over { [n]! [m-n]! }}.
\ee

\nd Let us mention that here we consider the $su_q(n)$, only at the algebra
level, the rest of its Hopf algebra structure is not needed for our
present purposes, but they remain worth studying from the physical point
 of view. 
 By analogy with the bosonization of the $su(n)$ by a set of canonical 
oscillators, the $q$-bosonization of the quantum $su(n)$ is always possible by
using the $q$-deformed oscillator of eq.(\ref{qosc}). Indeed for a set of $n$
such oscillators the bosonized generators are \cite{Hay}
\bea
e_i &=& a_i^{\dagger} a_{i+1} \;, \nn \\
f_i &=& a_{i+1}^{\dagger} a_{i} \;, \nn \\
k_i &=& q^{1/2 (N_i - N_{i+1})}.
\eea

\nd We also introduce the generators
$C_i = q^{1/2 (N_i + N_{i+1})}$, $i=1,2,...,n-1$,
which commute with any $\left \{ e_i,f_i,k_i \right \}$ with the same index.
Using this commutativity property and eq.(\ref{trans}) we express the 
$su_q(n)$ generators in terms of the QAL-bosons {\it i.e.}
\bea
e_i &=& b_i^{\dagger} b_{i+1} C_i q^{-1} \;, \nn \\
f_i &=& b_{i+1}^{\dagger} b_{i} C_i q^{-1}\;, \nn 
\eea

\nd By means of these equations and the relations (\ref{numb}) we now write the 
model's Hamiltonian as 
\be
H _{AL}= -\sum _{i=1} ^{n} \left \{ q C _i ^{-1} (e_i + f_i ) 
         -2 N_i \right \} ,
\ee

\nd Due to eq.(\ref{nu}), we can finally embed this Hamiltonian in the $su_q(n)$
algebra as
\be 
H_{AL}= -\sum _{i=1} ^{n} \left \{ q C_i^{-1} (e_i +f_i ) 
        -\sum _{j=1} ^{n-1} (\Omega ^{-1} )_{ij} h_i -(\Omega ^{-1} )_{in} h
        \right \}
\ee

\nd where
\be
C_i ^{-1} =q^{-{1\over{2}}(N_i +N_{i+1} ) } =
     q^{-{1\over{2}} \left \{
                     {1\over{2}} \sum _{j=1} ^{n-1} (\Omega ^{-1} ) _{ij} h_i
                                  +(\Omega ^{-1} ) _{in} h +
                     {1\over{2}} \sum _{j=1} ^{n-1} (\Omega ^{-1} ) _{i+1,j} 
                                 h_{i+1}+(\Omega ^{-1} ) _{i+1,n} h
                     \right\}
        }.
\ee

It is important at this point to emphasize that as our analysis
shows that the quantum group symmetry of the QAL model stems 
from the non-canonical character of the quantum variables defined at each 
site, (\cf eq.(\ref{nonCCR})). This non-canonical character of the degrees of
freedom is in fact necessary in order to prove the Hamiltonian structure
 of the AL equations with respect to the Hamiltonian of 
eq.(\ref{hAL}), \cite{Ger}. We conclude therefore 
that the found quantum group symmetry
is a genuine feature of the model~\cite{demos}.

 Similarly to the preceding case of the DNLS $su(n)$ invariances, the
 $su_{q}(n)$ dynamical symmetry of the QAL chain model implies that
 all quantum Casimir elements of $su_{q}(n)$ commute with $H_{QAL}$.
 In this way they become constants of the motion generated by the
 QAL Hamiltonian. For the construction of $q$-invariants of $su_{q}(n)$
 there are several approaches \cite{zhang,chak,bincer}. For our
 purpose a construction similar to that outline before for the
 DNLS case will be employed~\cite{bincer}. Let us introduce in terms
 of the $q$-deformed algebra generators the following elements:
 \bea
 h_{a} &\equiv& \e_{a}-\e_{a +1} \;\; \e_{1}+\e_{2}+\cdots +\e_{n}=0,\nn \\
 E_{ab} &\equiv& 
 \left\{ 
 \begin{array}{ll}
 E_{ac}E_{cb}- q^{-1}E_{cb}E_{ac} & a< c< b\;, {\rm no \ summation }\\
 e_{a} & a+1=b,\\
 (q-q^{-1})& a=b\;,
 \end{array} 
 \right. 
 \eea

 \nd and $E_{ab}=0$ for $a>b$. Also
 $F_{ba}\equiv E_{ab}^\dagger$,
 $e_a ^\dagger = e_a$, and the range of values for all indices is as
 before.

 \nd Notice that the $q$-deformed generators $E_{ab}(q)$, are not
 invariant under $q\rightarrow q^{-1}$, so we can define another
 set of generators $\tilde{E}_{ab}=E_{ab}(q^{-1})$, $\tilde{F}_{ab}=
 F_{ab}(q^{-1})$. Then we introduce the elements
 \be
 M_{ab}\equiv E_{ab}q^{e_a -2a}F_{cb}q^{e_c +2c},
 \ee

 \nd which yields the $q$-Casimir elements
 \be
 C_{2p}\equiv q^{2a}(M^p )_{aa}.
 \ee

 \nd By the exchange $q\rightarrow q^{-1}$, we obtain a new
 set of invariants
 \be
 \tilde{C}_{2p}\equiv C_{2p}(q^{-1})=q^{-2a}(\tilde{M^p})_{aa}\;,
 \ee

 \nd where $\tilde{M^{p+1}}$ as before is obtained recursively from
 the operator-valued matrix
 \be
 \tilde{M}_{ab}=M_{ab}(q^{-1})=\tilde{E}_{ac}q^{- e_a +2a}
 \tilde{F}_{ab}q^{- e_c -2c}\;.
 \ee

 Finally two Casimir generators invariants under the exchange
 $q\rightarrow q^{-1}$
 may be constructed by combining the two previous ones
 in the form
 \bea
 {\cal C}_{2p}&=&(C_{2p}+\tilde{C}_{2p}) / (q+q^{-1}) \nn \\
 {\cal C}_{2p+1}&=&(C_{2p}-\tilde{C}_{2p}) / (q-q^{-1}).
 \eea

\nd These last expressions of the invariant Casimir operators modified
 with appropriate coefficients reduce presicely in the $q\rightarrow 1$
 limit to the Casimir operators of equal degree for the non-deformed
 $su(n)$ algebra (see \cite{bincer}, for details and some additional expressions
 for $q$-Casimir elements).

If we now confine ourselves to a two-site restriction of the QAL-chain then
the 
Hamiltonian 
\be
\label{QAL}
H_{AL}=-qC_1 ^{-1} (a_1 ^\dagger a_2 +a_2 ^\dagger a_1 ) +
        2(N_1 +N_2),
\ee
has a $su_q(2)$ dynamical symmetry. This symmetry will be utilized in the next 
section in order to solve the eigenvalue problem of the AL-dimer.
 
\section{Exact solutions}
\subsection{\it The DNLS dimer}
We now turn to the analytic determination of the eigenvalues and eigenvectors
of the quantum DNLS chain in the simplest case of two sites. As was already
mentioned there is a
number of previous investigations in the case of classical
dimer, which predict the well known phenomenon of selftrapping~\cite{KC}.
But also for the quantum case there are numerical and perturbative 
efforts \cite{BES}, and quasiclassical treatments~\cite{EJC} along with
formal techniques based on the inverse
scattering method~\cite{ism}.
  aiming to study the quantum manifestations
of selftrapping, 

Our approach will be entirely analytic and will be based in the $su(2)$ 
pseudo-spin expression given to $H_{QD}$ Hamiltonian (\ref{hqd}).
According to previously given general prescription  of the group 
symmetry of the chain, the two-boson realization of the $su(2)$ generators reads

\bea
J_+ &=& a_1 a_2 ^\dagger ,\nn \\
J_- &=& a_1^\dagger a_2 , \nn \\
J_0 &=& {1\over{2}} (N_1-N_2).
\eea

\nd The generators obey the standard relations of the quantum angular momentum
\bea
[J_0 , J_{\pm}] & = & \pm J_{\pm} , \nn \\
\left[J_+ , J_- \right]   & = & 2 J_0 .
\eea

\nd With the Casimir operator adapted from the preceding general formulae
\be
C=J_0 (J_0 -1)+J_+ J_- = J_0 (J_0 +1)+J_- J_+ \;,
\ee

\nd the matrix form of the Hamiltonian is induced by the $(2j+1)$-dimensional
matrix representation of the generators in the spin basis 
$\left \{ |jm> \right \} _{m=-j} ^{j} $ 
where $j={1\over{2}},1,{3\over{2}},...$, reads:
\bea
C|jm>&=&j(j+1)|jm>, \nn \\
J_0 |jm>&=&m |jm>,\nn \\
J_\pm |jm>&=&\sqrt{(j\mp m)(j\pm m+1)}|j m\pm>\;.
\eea

\nd  In the two boson
labeling of the spin state vector 
\be
|j={1\over{2}} (n_1 + n_2 )\ ;\ m={1\over{2}} (n_1-n_2)>\equiv |n_1,n_2>,\ \ 
n_{1,2}\geq 0 ,
\label{x}
\ee

\nd we see that the total boson excitation number $n_1+n_2$, determines the 
dimensionality of the Hamiltonian matrix, and also the total energy 
allocated to the dimer. Then the eigenvalue equation
\be
H_{QD} |\phi >=\lambda |\phi >
\label{eigen}
\ee

\nd for a Hamiltonian eigenvector which is expressed by
\be
|\phi >= \sum _{n=-j} ^{j} c_n |n>,
\label{expa}
\ee

\nd leads to the difference equation
\be
\gamma n^2 c_n +\epsilon \left( \sqrt{(j+n)(j-n+1)} c_{n-1} 
+\sqrt{(j-n)(j+n+1)} c_{n+1} \right )=\lambda c_n .
\ee

\nd with boundary conditions $c_{-j-1}=c_{j+1}=0$.
We introduce new variables $\psi _n=c_n \gamma _n$ and impose upon  
$\gamma _n $'s the relation
\be
\gamma _n \sqrt{ (j+n)(j-n+1) } =\gamma _{n-1}
\ee

\nd with solution 
\be
\gamma _{-j+n}=\sqrt{ n!\prod _{i=0} ^{n-1}(2j-i) }\gamma _{-j}
\ ,\ \ \ n=1,2,...,2j \;.
\ee

\nd In this way we  obtain a normalized form of the iteration
\be
\psi _{i-j+1}= (\lambda +\mu _i ^{(1)})\psi _{i-j} +\mu _i ^{(2)}
 \psi _{i-j-1}
\ \ ,\ \ \ i=0,1,...,j ,
\label{rer}
\ee

\nd where 
\be
\mu _i ^{(1)}=\gamma (i-j) ^2 \nn
\ee
\nd and
\be
\mu _i ^{(2)}=-i(2j+1-i).
\ee

\nd (also we set $\mu _i ^{(0)}=1$ for later use).
Then from the boundary conditions
$\psi _{-j-1} =\psi _{j+1}=0$ and the seed of the recurrence relation
$\psi_{-j}=1$ ,
we deduce that the polynomial $p(\lambda )\equiv \psi _{j+1} (\lambda )$,
is the characteristic polynomial of the Hamiltonian matrix $H_{QD}$. Let 
us write 
\be
p(\lambda )= \sum _{i=0} ^{2j+1} \d _i ^{(2j)} \lambda ^2 ,
\ee

\nd then the coefficients of the characteristic polynomial are determined by 
induction. Indeed if 
we introduce the concept of the weight of a coefficient  $\d _i ^{(k)}$
defined as
\be
{\rm wt} (\d _i ^{(k)})\equiv w_i={\rm deg }p(\lambda )-i =2j+1-i ,
\ee

\nd then we shall obtain the following expression for the coefficients

$$
\d _i ^{(\nu )}= \sum _{i_1,i_2,...,i_{w_i}=0} ^2
  \ \  \sum _{l_{w_i}=i_1+...+i_{w_i}-1} ^\nu
  \ \   \sum _{l_{w_{i-1}}=i_1+...+i_{w_{i-1}}-1} ^{l_{w_i}-i_{w_i}}
    ...\\
$$
\be
  \ \   \sum _{l_2=i_1+i_2-1} ^{l_3-i_3}
  \ \  \sum _{l_1=i_1-1} ^{l_2-i_2}
  \ \   \mu _{l_{w_i}} ^{(i_{w_i})}...
    \mu _{l_2} ^{(i_2)}
    \mu _{l_1} ^{(i_1)} ,
\label{coef}
\ee

\nd where the summation indices satisfy the constrains
\be
i_1 \neq 0,\ \ \ i_{w_i} \neq 2,\ \ \  i_1+i_2+...i_{w_i} = w_i ,
\ee
\nd and if $  i_k=2$ then $ i_{k+1}=0$.

Next we determine the associated  eigenvectors by specifying
their expansion coefficients in the spin basis (\cf eq.(\ref{expa})).
Since the $n$-th of these coefficients is identified with the $n$-th order
iteration of the recurrence relation issued in eq.(\ref{rer}) we write
\be
\psi _{-j+n} (\lambda _a ) = q(\lambda _a ).
\ee

\nd The polynomial $q(\lambda _a )$ for each of the eigenvalues
$\left\{\lambda _a \right \} _{a=0} ^{2j} $,
has deg$q(\lambda _a) =n$, then we can write 
\be
q(\lambda _a)=\sum _{i=0} ^n \d _i ^{(n-1)} \lambda _a ^i .
\ee

\nd The finding of the eigenvector amounts then to the determination of
the coefficients $\d _i ^{(n-1)}$. This is done by use of eq.(\ref{rer})
and the solution is again given by eq.(\ref{coef}), with $\d _i ^{(-1)}=1$.
The variable $w _i $ this time stands for the weight of the coefficients of the 
$q(\lambda )$ polynomial and it is defined by 
\be
{\rm wt}(\d _i ^{(n-1)})=w_i=n-i\ ,\ \ \ i=0,1,...,n \;.
\ee

Finally, we shall show that the found operator spectrum possesses
proper mathematical properties, namely that the 
energy eigenvalues are real numbers
with zero multiplicity (non-degenerate) and that the corresponding set of
eigenvectors form an orthonormal and complete basis. To this end we 
first recall
some generalities~\cite{Sze}. Let us assume that  we have a set of single variable polynomials 
$\{q_n (x)\}_{n=0}^{N-1}$, obeying the three term recurrence relation
\be
\a_n q_{n+1}(x) + \b_n q_n (x) + \g_n (x) = x q_n (x) \;,
\ee

\nd with $q_{-1} (x)=0$, which have roots $x_i$, \ie $p_n (x_i ) =0$, if we
introduce the the numbers $d_n = \sqrt{\frac{\g_n}{\a_{n-1}}}d_{n-1}$,
with $d_0$ arbitrary positive number and the variables $a_n = \a_n a_{n+1}$,
then the socalled  Darboux-Christoffel (DF) formula given by
\be
\sum_{n=0}^{N-1} \frac{q_n (x) q_n (y)}{d_n ^2} = \frac{a_{N-1}}{a_N}
\frac{1}{d^{2}_{N-1}}\frac{q_{N} (x) q_{N-1} (y)  - q_{N-1} (x)  q_{N} (y) }
{x-y}\;,
\ee

\nd is valid for the polynomials. For two roots $x=x_i , y=x_j$ the DF
formula becomes
\be
\sum_{n=0}^{N-1} \frac{q_n (x) q_n (y)}{d_n ^2} = \d_{ij} {\cal N}^2 _i
\label{dff}
\ee

\nd with 
\be
{\cal N}^2 _i = \sum_{n=0}^{N-1} \frac{q_n ^2 (x_i)}{d_n ^2} = \frac{a_{N-1}}{a_N}
\frac{1}{d^{2}_{N-1}} q^{'}_{N}(x_i) q_{N-1}(x_i)\;.
\ee

\nd The non-degeneracy of the roots is now obtained from eq.(\ref{dff}), which
implies that for $x_i = x_j$, we get $q^{'}_{N}(x_i) \neq 0$. 
Turning now to our case we can write for the $H_{DNLS}$ eigenvectors
the expression
\be
|\psi (\lambda _a ) >= {\cal N}_a \sum _{k=0} ^{2j}
     {\psi _{-j+k} (\lambda _a) \over {\epsilon _k }} |-j+k> ,
\ee

\nd where the normalization constant (which determines the value of $\g_{-j}$)
\be
{1 \over {{\cal N}_a}} = \sum _{k=0} ^{2j} 
    {\psi _{-j+k} ^2 (\lambda _a) \over {\epsilon _k ^2}},
\ee

\nd involves the factor
\be
\epsilon _k = \delta _{o,k} + k! \sqrt {
    \left ( 
    \begin{array}{c}
    2j\\
    k
    \end{array}
    \right )
    } \;,
\ee

\nd with $a=0,1,\ldots ,2j$.

\nd To verify the orthonormality of the eigenvectors we resort to the quoted
DF formula and write
\be
\inner{\psi(\l_a)}{\psi(\l_b)} = {\cal N}_a{\cal N}_b \sum _{k=0} ^{2j}
\frac{\psi _{-j+k} (\lambda _a) \psi _{-j+k} (\lambda _b)}
{\epsilon _k ^2 } =\d_{ab}\;.
\ee

\nd We note that the orthonormality of the eigenvectors is equivalent to the
orthonormality of the coefficients of the eigenvectors in the spin basis 
taken as polynomials with discrete variable, which is
the corresponding eigenvalue. Also the completeness of the eigenvalue basis
can be based on the DF formula; indeed it is straightforward to write:
\be
 \sum _{a=0} ^{2j} \proj{\psi(\l_a)}{\psi(\l_a)} = 
  \sum _{k,l =0} ^{2j} \left \{ \sum _{a=0} ^{2j}
{\cal N}^{2}_{a} \frac{\psi _{-j+k} (\lambda _a) \psi _{-j+k} (\lambda _a)}
{\epsilon_{k} \epsilon_{l}} 
\right \} 
\proj{-j+k}{-j+l} = {\bf 1}\;.
\ee

Having constructed the eigenvector basis of the Hamiltonian, we can now
investigate dynamical questions, by expanding a suitable initial state vector on 
this basis, here however we refrain from doing so and shortly we will
turn to the
study of the quantum manifestations of the classical selftrapping.

\subsection{\it The QAL dimer.}

\nd In this section we  put forward an analysis of the eigenvalue 
problem for the QAL-dimer based, as in the DNLS-dimer, on the symmetry
algebra of the $H_{QAL}$ Hamiltonian. The new feature in this case is that we 
deal with the quantum, or $q$-deformed algebra $su_q(2)$. As the value of 
$q$-deformed
parameter is a non-negative real number, the representation theory (which as we 
have shown in the preceding case is an indispensable element for the concise 
formulation of the eigenvalue problem) of $su_q(2)$ is very similar 
to its $q=1$ limiting case.

Starting with the Hamiltonian of eq.(\ref{QAL}) and after appropriate shifting 
and scaling with constant factors we get 
\be
H_{AL}= \sum _{i=1} ^{2} a_i^\dagger (a_{i+1} +a_{i-1}).
\ee
\nd This can be expressed in terms 
of the effective quantum $su(2)$ generator ( \viz
\be
H_{AL}=J_+^q+J_-^q .
\ee

\nd (We note here that the $H_{AL}$ Hamiltonian is exactly that of 
Azbel-Hofstadter model which describes Bloch electrons in 
magnetic field~\cite{qqq}, however in that case the $q$-deformation parameter
is determined by the value of the magnetic flux and is root of unity;
given that the latter is solvable by Bethe ansatz techniques the relation between 
the two models is worth studying.) 
These quantum generators bosonized in an angular momentum fashion
become
\bea
J_+ ^q &=& a_1 a_2 ^\dagger , \nn \\
J_- ^q &=& a_1 ^\dagger a_2 , \nn \\
J_0 ^q &=& {1\over{2}} (N_1 -N_2)\;. 
\eea

\nd They satisfy the quantum $su_q(2)$ algebra commutation relations \ie
($q \equiv e^s$ ),
\bea
[ J_0 ^q ,J_\pm ^q ] &=& \pm J_\pm ^q \nn \\
\left[J_+ ^q , J_- ^q \right] &=& [ 2J_0 ^ q ]
={\sinh (2sJ_0^q) \over { \sinh s }}
\eea

\nd With the $q$-Casimir operator adapted from the preceding general formulae
for $q$-deformed central elements,
\be
C_q =[J_0^q ][ J_0^q -1]+J_+^q J_-^q = [J_0^q ]  [J_0^q +1]+J_-^q J_+^q ,
\ee

\nd the matrix representation of the q-deformed generators
\bea 
C_q |jm> &=& [j][j+1]|jm>, \nn \\ 
J_ \pm ^q |j,m> &=& \sqrt { [j\mp m][j \mp m+1 ] } |j,m+1> ,\nn \\
J_0^q |j,m>&=& m|j,m> ,
\label{reppm}
\eea

\nd is very similar to their $q=1$ counterparts. We notice that the 
non-linearity of the model resites in the fact that the off-diagonal matrix
elements of the Hamiltonian matrix are expressed in terms of  $q$-numbers
({\cf eq.(\ref{reppm})).
Simple inspection shows that in the limit $\gamma \rightarrow 0$,
$q \rightarrow 1$, where for any $q$-number the limit $[x]\rightarrow x$, is valid, the 
generator matrices $J_\pm ^q$ become the $su(2)$ generators
\ie $J_\pm ^q \rightarrow  J_\pm $. 
In this limit the QAL-dimer becomes a linear angular momentum problem that is
easily solved.

To address the eigenvalue problem we first note that a similar relation 
as in eq.(\ref{x}) for $su(2)$, holds between the state vectors of
$su_q(2)$ and their parametrization in terms of the excitation numbers
of two $q$-bosons. If we  now
assume an expansion $|\phi >=\sum _{n=-j} ^j c_n |n> $, for the 
eigenvector $|\phi >$, associated to the eigenvalues $\lambda $ 
\ie
$H_{QAL} |\phi >=\lambda |\phi >$, then this last relation by means of 
eqs.(\ref{hAL}),(\ref{reppm}) yields
\be
\sqrt { [j-n+1][j+n] } c_{n-1} + \sqrt { [j+n+1][j-n] } c_{n-1} = \lambda c_n .
\ee

\nd Again we normalize this difference equation with a change of variables 
$c_n= { \psi _n \over { \gamma _n}} $, where the $\gamma _n$'s satisfy the 
relation
\be 
\gamma _n =\gamma _{n-1} \sqrt{ [j+n][j-n+1] }   ,
\ee

\nd and explicitly are given by 
\be
\gamma _{-j+n}= \sqrt{ [n]! \prod _{i=0}^{n-1} } \gamma _{-j}\ \ , \ \ 
n=1,2,\ ...\ ,2j\;.
\ee

\nd  Then the normalized iteration  with 
vanishing boundaries \ie $\psi _{j+1}=\psi _{-j-1}=0$ and seed $\psi_{-j}=1$
reads,

\be
\psi _{n+1} = \lambda \psi _n - [j+n][j-n+1]\psi _{n-1} .
\label{qiter}
\ee

\nd As was explained in the similar situation of the DNLS-dimer, the
eigenvalues are the roots of the real polynomial $\psi _{j+1}(\lambda )=0.$
By induction we find that one needs to distinguish between integer and 
half-integer values for the index $j$. In other words we need to distinguish 
between even total number of excitation quanta and odd total number of 
excitation quanta that are available in the two $q$-boson sites of the model,
(\cf eq.(\ref{x}). Explicitly we find that for $j$-integer  
\be
\psi _{j+1} (\lambda ) = \lambda q(\lambda ^2 ) =0
\ee

\nd where deg$q(\lambda )=j$ and deg$\psi _{j+1}(\lambda )=2j+1 $.
This implies that the energy spectrum contains the zero eigenvalue and that
the remaining energy levels are arranged in doublets for each 
pair $\pm \lambda $. If we now define
 $\chi = \lambda ^2 $, and assume that
\be
q(\chi ) = \sum _{m=0} ^ j \beta _m ^{(2J)} \chi ^m ,
\ee
then we seek to determine the  coefficients $\beta _m ^{(2J)}$.

\nd Similarly for $j$-half-integer we obtain that 
\be
\psi _{j+1} (\lambda ) = p(\lambda ^2 ) =0 ,
\ee

\nd where deg$p(\lambda )={2j +1 \over {2}}$, and
\be
p(x)=\sum _{m=0} ^{ {1\over{2}}(2j+1) }\beta _m ^{(2J)} \chi ^m .
\ee

\nd Then by induction of the recurrence relation of eq.(\ref{qiter}), we 
specify the $(j+1)$-order of iteration $\psi _{j+1} (\lambda )$, which provides
the coefficients
\be
\beta _m ^{(\nu )}= \sum _{i_{j-m} = p _{j-m} } ^{\nu }
        \ \            \sum _{i_{j-m-1} = p _{j-m-1 } } ^{i_{j-m}-2 }
		    ...
                    \sum _{i_2 = p _2} ^{i_3-2 }
           \ \         \sum _{i_1 = p _1} ^{i_2-2 }
		    \mu _{i _{j-m}}
		    ...
		    \mu _{i _2}
		    \mu _{i _1}.
\label{qsol}
\ee

\nd In this expression we have used the abbreviation 
\be
\mu _n =-[j+n][j-n+1],
\ee

\nd and the $p_r$'s are indices determined by the iteration 
\be
p_r=p_{r-1} +2  \ \ \ r=2,3,...\ \ ,\ \ p_1=1.
\ee

\nd Next we construct the set of $(2j+1)$-eigenvectors of $H_{QAL}$
by specifying their expansion coefficients $\psi _{-j+n}$, $n=0,1,...,2j$, 
in the $q$-spin basis. We find that for $n$-even 
\be
\psi _{-j+n} (\lambda )= u(\lambda ^2 ),
\ee
and
\be
u(\chi )=\sum _{m=0} ^{n/2} \beta _m ^{(n-1)} \chi ^m .
\ee

\nd Similarly for $n$-odd
\be
\psi _{-j+n} (\lambda )= w(\lambda ^2 ),
\ee
with
\be
w(\chi )=\sum _{m=0} ^{1/2(n-1)} \beta _m ^{(n-1)} \chi ^m .
\ee

\nd where the coefficients $\delta _m ^{(n-1)}$ are determined again by the 
relation (\ref{qsol}).

Concerning normalization of the eigenvectors we note that since we have set
$\psi _{-j}=1$, the remaining unspecified variable $\gamma _{-j}$,
 is determined from the normalization condition of the 
eigenvectors as in the preceding case. The latter are given explicitly by   
\be
|\psi (\lambda _a ) >= {\cal N}_a \sum _{k=0} ^{2j}
     {\psi _{-j+k} (\lambda _a) \over {\epsilon _k }} |-j+k> ,
\ee

\nd where the normalization constant
\be
{1 \over {{\cal N}_a}} = \sum _{k=0} ^{2j} 
    {\psi _{-j+k} ^2 (\lambda _a) \over {\epsilon _k ^2}},
\ee

\nd involves the factor
\be
\epsilon _k = \delta _{o,k} + [k]! \sqrt {
    \left [ 
    \begin{array}{c}
    2j\\
    k
    \end{array}
    \right ] _q
    } ,
\ee

\nd for $a=0,1,...,2j$.
 
\section{ Numerical Results and Discussion}

We  use now the previously derived exact results to study  the
spectral behavior of QDNLS and QAL models.  These results
are shown in Figure 1 for QAL and in Figure 2 for QDNLS respectively.  
In Figure 1a we present the spectrum
of QAL as a function of nonlinearity for $j=4$. We note the distinct
"repulsion" of the levels:  the larger the value of nonlinearity the greater
the repulsion between the levels.  This tendency for repulsion of the energy
levels is not related to similar repulsive and avoiding crossing behavior
 one encounters in quantum non-integrable systems (see \eg~\cite{haake}), since our 
 QAL-dimer is an integrable model. It should rather be attributed to 
 the non-linear dependence of the eigenvalues on the $q$-deformation 
 parameter, that is to say that it should be an effect of the quantum group 
 symmetry itself.
  In Figure 1b
we plot the lowest QAL energy levels for a given 
nonlinearity value ($\gamma = 2$)
but as a function of the number of quanta in the system.  The classical AL
is obtained in the limit of very large $j$-values.  We note here the repulsive
aspect of the levels as well.  

In Figure 2a  the spectrum of QDNLS 
is shown for $j=3$ as a function of the DNLS nonlinearity parameter.  We note
a great difference from the corresponding spectrum of QAL; here pairs of adjacent
levels are grouped together and they merge as nonlinearity
increases.  This property is evident for all pairs of levels and especially
in the lowest ones.  We note that the rate of convergence is $\gamma$-dependent.
In order to investigate this property we plot in Figure 2b the energy difference
of the lowest two pairs of levels as a function of the nonlinearity parameter in
a log-log plot.  The labels $(1)$ and $(2)$ signify the lowest and
next to lowest pairs respectively.  There is a drastic change in the 
convergence plot as the amount of nonlinearity exceeds a certain value.
 While at small $\gamma$ the energy difference shrinks very slowly,
the rate changes very rapidly for large $\gamma$.  The convergence is done
seemingly in an algebraic fashion with energy dependent exponents that can be
easily obtained. 

The numerical results obtained for the energy splittings demonstrate
the role of classical selftrapping in the quantum mechanical regime.
In its absence, as in the AL equation, the system energy spectrum shows
repulsion, while in its presence the opposite effect is manifested, \viz level
clustering.  This energy clustering in the quantum regime is a 
signature of long-lived excitations.  Indeed, if an excitation is created on
one site of the QDNLS dimer, while the latter is in a large nonlinearity
regime, the tunneling time will be very large, leading thus to a very long lived
and localized excitation.  This excitation can be thought of as a quantum breather
induced by the nonlinearity of the corresponding  classical problem. Such a 
breather does not exist in the QAL case, as becomes evident from its spectrum. 
Both classical counterparts of the two problems studied in this work 
are integrable;  it would be interesting to investigate in detail and
compare cases with and without selftrapping but while the system is also
non integrable.  In such problems the simultaneous presence of 
classical localization and chaos will have interesting reprecautions in
the quantum mechanical spectrum.

\section{Acknowledgments}
We are grateful to G. P. Tsironis for constructive discussions. Also
we acknowledge support from the Greek 
Secretariat of Research and Technology under contract
$\Pi ENE\Delta$ 95/1981

\pagebreak
\centerline{\bf Figure captions}

\nd Fig. 1a. The spectrum
of QAL as a function of nonlinearity for $j=4$, \ie nine energy quanta.

\nd Fig. 1b. The lowest QAL energy levels for a given 
nonlinearity value ($\gamma = 2$)
as  function of the number of quanta in the system.

\nd Fig. 2a. The spectrum of QDNLS 
is shown for $j=3$ as a function of the DNLS nonlinearity parameter.

\nd Fig. 2b. The difference
of the lowest two pairs of energy levels 
as a function of the nonlinearity parameter in
a log-log plot.  The labels $(1)$ and $(2)$ signify the lowest and
next to lowest pairs respectively.


\newpage
\begin{thebibliography}{25}

\bibitem{KC}
V. M. Kenkre and D. K. Campbell, Phys. Rev. B {\bf 34}, 4959 (1986);\\
G. P. Tsironis and V. M. Kenkre, Phys. Lett. A {\bf 127}, 209 (1988);\\
V.M. Kenkre, G.P. Tsironis and D.K. Campbell, in "Nonlinearity in condenced 
matter" (Edited by R. Bishop et al.) (Springer 1987);\\
V.M. Kenkre, Physica {\bf D68}, 153 (1993).


\bibitem{ELS}
J.C. Eilbeck, P.S. Lomdahl, and A.C. Scott, Physica (Amsterdam) {\bf 16D}, 318 
(1985).

\bibitem{BES}
L. Bernstein, J.C. Eilbeck, and A.C. Scott, Nonlinearity {\bf 3}, 
293 (1990);\\
L. Bernstein, Physica (Amsterdam) {\bf 68D},174 (1993);\\
J C Eilbeck and A C Scott, {\it Quantum lattices}, in
``Nonlinear Coherent structures in Physics and Biology", NATO ASI
Series B: Physics {\bf 329}, Eds. K. H. Spatschek and F. G. Mertens,  Plenum
Press, 1-14 (1994).\\
S. Aubry, S. Flach, K. Kladko and E. Olbrich, Phys. Rev. Lett.
{\bf 76}, 1607 (1996).

\bibitem{AL}
M.J. Ablowitz, J.F. Ladik, Stud. Appl. Math. {\bf 55}, 213 (1976).

\bibitem{ism}
V. Z. Enol'skii, M. Salerno, A. C. Scott and J. C. Eilbeck, 
Physica (Amsterdam) {\bf 59D},1 (1992);\\
V. Z. Enol'skii, V. B. Kuznetzov and M. Salerno, 
Physica (Amsterdam) {\bf 68D},138 (1993).

\bibitem{Ger}
V.S. Gerdjicov, M.I. Ivanov and P.P. Kulish, J. Math. Phys. {\bf 25} 
 25 (1974).

\bibitem{EJC}
D. Ellinas, M. Johansson and P. L. Christiansen, 
"Quantum nonlinear lattices and coherent state vectors", 
Physica D, in press 1999.

\bibitem{CP}
M. Chaichian and A. Demichev, {\it Introduction to Quantum Groups} (World 
Scientific, 1996).\\
V. Chari and A. Pressley, {\it A Guide to Quantum Groups}, (Cambridge
University Press, 1994).

\bibitem{Mac}
A.J. MacFarlane, J. Phys. {\bf A22} 4581 (1989);\\
L.C. Biedencharn, J. Phys. {\bf A22} (1989) L871;\\
M. Chaichian and P.P. Kulish, Phys. Lett. {\bf B234} (1990) 72;\\
P.P. Kulish and E.V. Damaskinsky, J. Phys. A:Math. Gen  {\bf 23} 
L415 (1990);\\
M. Chaichian, D. Ellinas  and P.P. Kulish, Phys. Rev. Lett. 
{\bf 65} 980 (1990).

\bibitem{Sal}
M. Salerno, Phys. Rev. A {\bf 46},6856 (1992);
Phys. Lett. A {\bf 162}, 381(1992).

\bibitem{W1}
P.A.M. Dirac, {\it The principles of quantum mechanics}, (Oxford University 
Press, London, 1947);\\
H. Weyl, {\it The theory of groups and quantum mechanics.} Dover Publications,
New York, 1950. (Published by E.P. Dutton and Company in 1931.)

\bibitem{Gil}
R. Gilmore, {\it Lie groups, Lie algebra and some of their applications} 
(J. Wiley \& Sons, N. York 1974).

\bibitem{barut}
A. Barut and R. Raczka, {\it Theory of Group Representations and 
Applications} (2nd Ed. World Scientific Singapore 1986);\\
{\it Dynamical Groups and Spectrum Generating Algebras} eds. 
A. Barut, A. Bohm, Y. Ne'eman (World Scientific Singapore 1989).

\bibitem{demkov}
D. Ellinas and V. Kovanis,
Phys. Rev. A {\bf 51}, 423 (1995).
 
\bibitem{Kulish}
P. P. Kulish, Theor. Math. Phys. {\bf 86} 108 (1991);\\
G. Rideau, Lett. Math. Phys. {\bf 24} 147 (1992);\\
M. Chaichian, H. Grosse and P. Pre\v{s}najder, J. Phys. {\bf A 27} 2045 (1994).

\bibitem{Ext}
H. Exton, {\it q-Hypergeometric Functions and Applications }
(Horwood, Chichester, 1983)

\bibitem{Hay}
T. Hayashi, Commun. Math. Phys. {\bf 127}, 129 (1990).

\bibitem{demos}
D. Ellinas, These proofs were announced in
talks given at:
"V Colloquium "Quantum groups and integrable systems" ", 
Prague,Czech Republic, June 1996;\\
Conference on: "Fluctuations, Nonlinearity and Disorder"
Heraklion, Crete Greece, Sept. 1996;\\
Workshop on Localization in nonlinear lattices,
Max-Planck-Institut f\"ur Komplexer Systeme, Dresden, Germany,
April 1997;\\
Needs '97, 11th Workshop on Nonlinear evolution
equations and dynamical systems, OAK, Kolymbari, Chania Crete Greece,
June 1997.

\bibitem{zhang}
R. B. Zhang, M. D. Gould and A. J. Bracken, J. Phys. A: Math. Gen. {\bf 24},
937 (1991);\\
Commun. Math. Phys. {\bf 137}, 13 (1991).

\bibitem{chak}
A. Chakrabarti, J. Math. Phys. {\bf 32}, 1227(1991).

\bibitem{bincer}
A. M. Bincer, J. Phys. A: Math. Gen. {\bf 24}, L1133 (1991).





\bibitem{Sze}
G. Szeg\"o, {\it Orthogonal polynomials}, AMS Colloq. Publ. {\bf 23}
(AMS Providence, Rhode Island 1975);\\
A. F. Nikiforov, S. K. Suslov, and V. B. Uvarov,
{\it Classical orthogonal polynomials of a discrete variable}, (Spinger-Verlag
Berlin 1991).

\bibitem{qqq}
P. B. Wiegmann and A. V. Zabrodin, Phys. Rev. Lett. {\bf 72}, 1890 (1994).\\
G. G. Athanasiu and E. G. Floratos, Phys. Lett. B {\bf 352}, 105 (1995).

\bibitem{haake}
F. Haake, {\it Quantum Signatures of Chaos}
 (Springer Series in Synergetics, Vol 54, Berlin 1991). 

                       
\end{thebibliography}
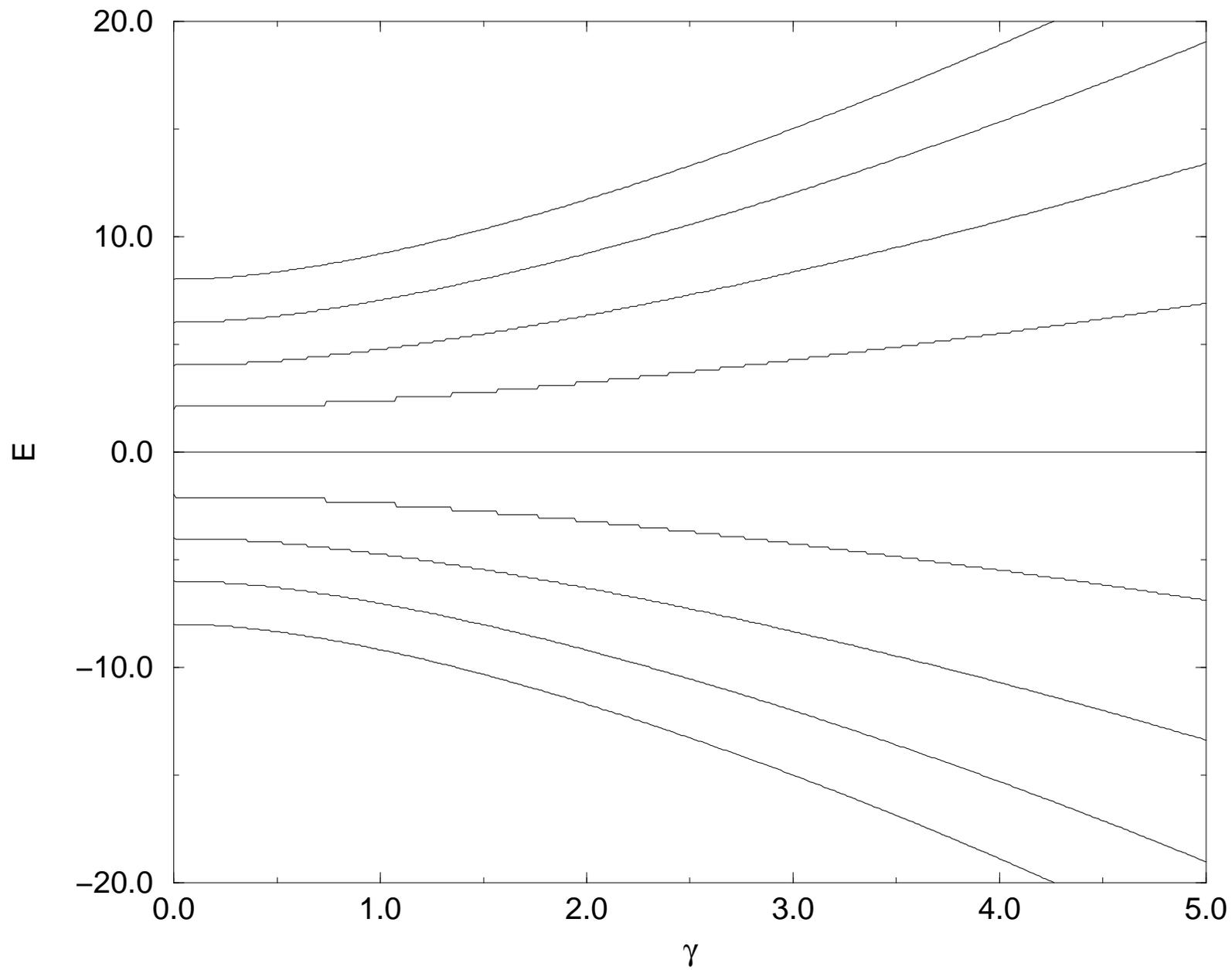
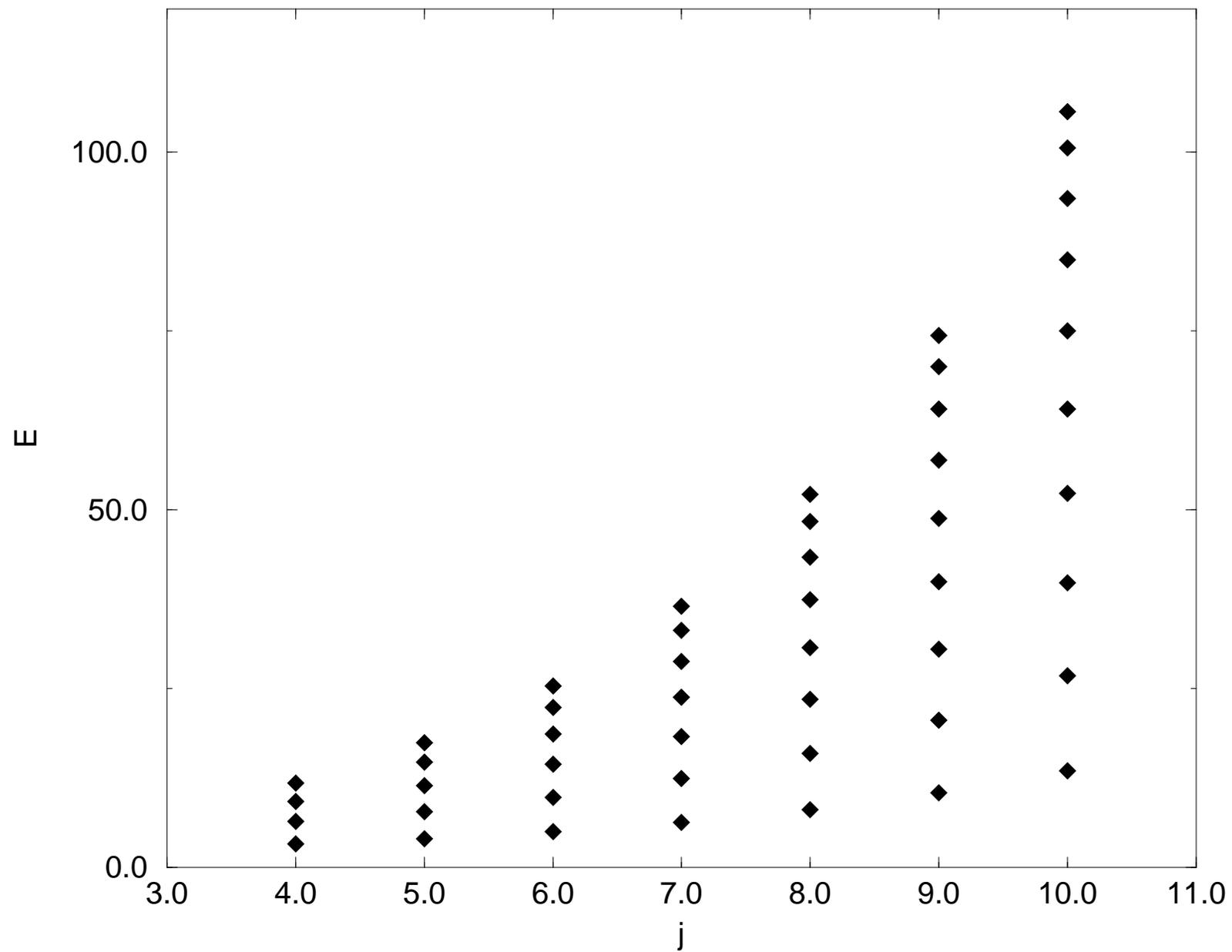
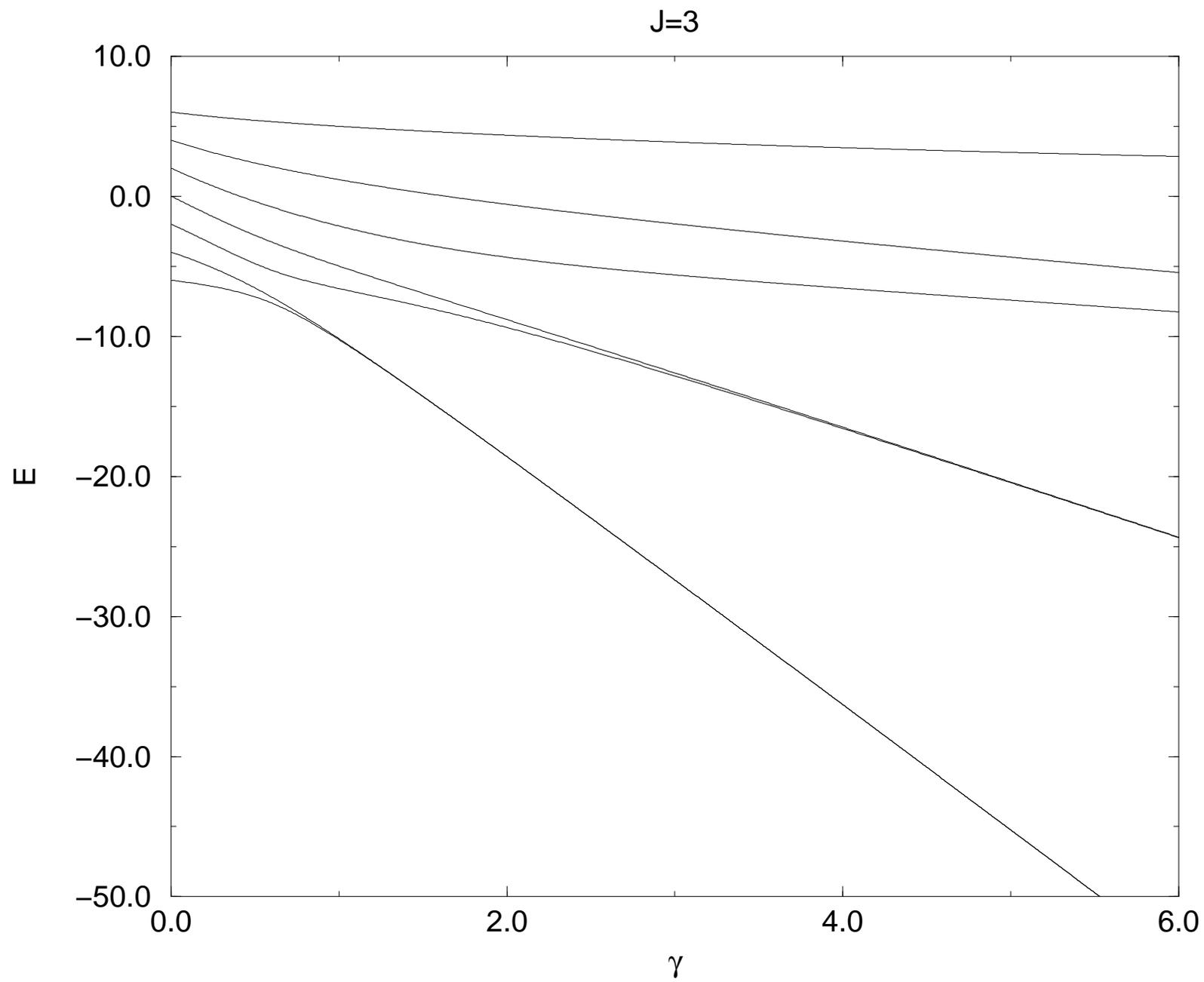
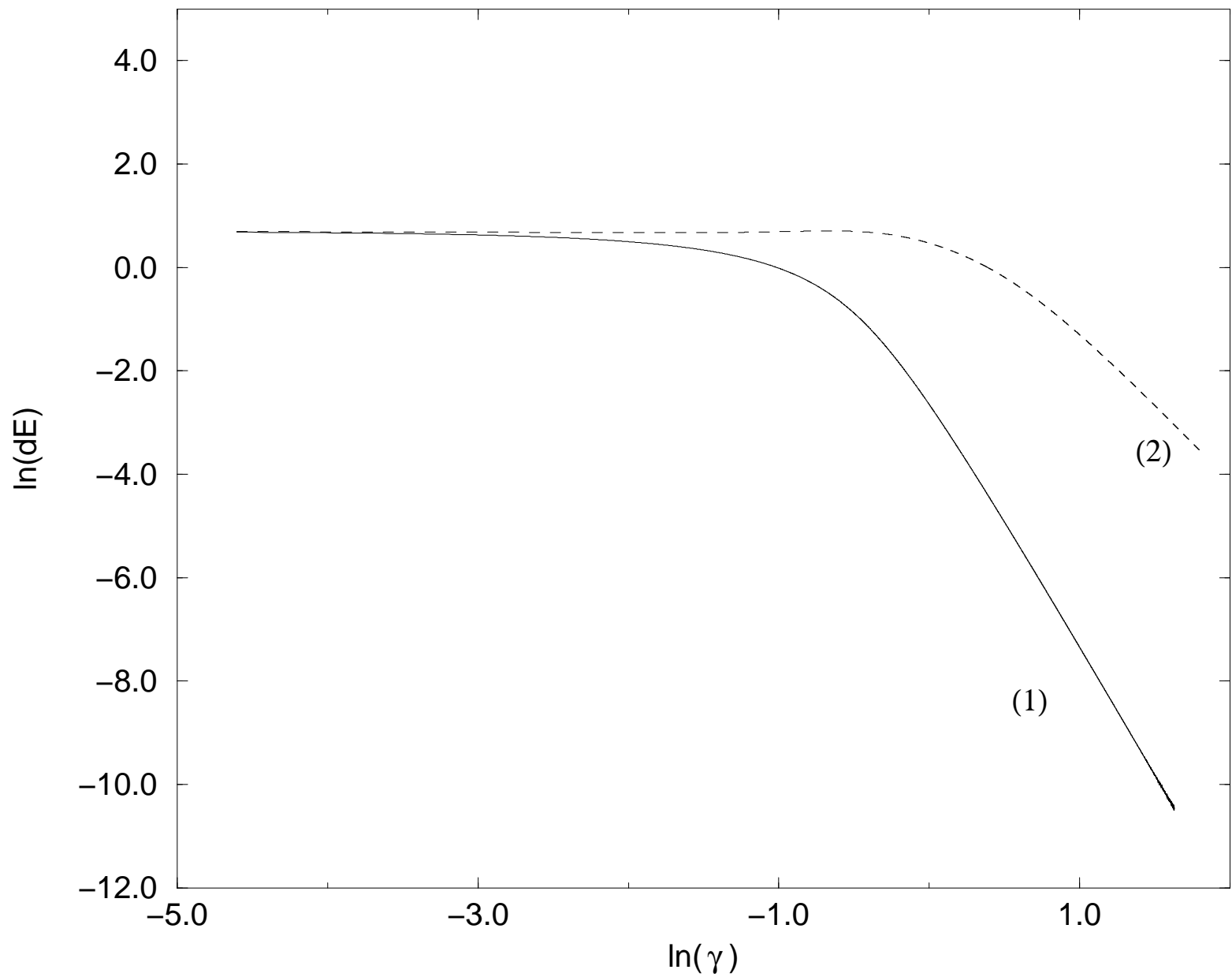
\end{document}